\documentclass{ws-procs9x6}

\def\hbar  {$\overline{\mathrm{H}}$}
\def\pbar  {$\overline{\mathrm{p}}$}

\def\pbarn  {{\pbar}N}
\def\pos    {$\mathrm{e}^{+}$}
\def\ele     {$\mathrm{e}^{-}$}
\def\na     {$^{22}\mathrm{Na}$}
\def\crystSize  {(13$\times$17.5$\times$17 mm$^3$)}
\def\others {{\it et al.}}

\begin{document}

\title{Detection of antihydrogen annihilations with a Si--micro--strip and pure CsI detector}
\author{I. Johnson$^1$, M. Amoretti$^2$, C. Amsler$^1$, G. Bazzano$^{3,4}$,
\\ G. Bonomi$^5$, A. Bouchta$^5$, P. Bowe$^6$, C. Carraro$^{2,7}$, C. L. Cesar$^8$,
\\ M. Charlton$^9$, M. Doser$^5$, V. Filippini$^{3,4}$, A. Fontana$^{3,4}$,
\\ M. C. Fujiwara$^{10,11}$, R. Funakoshi$^{11}$, P. Genova$^{3,4}$, J. S. Hangst$^6$,
\\ R. S. Hayano$^{11}$, L. V. J$\o$rgensen$^9$, A. Kellerbauer$^5$,
\\ V. Lagomarsino$^{2,7}$, R. Landua$^5$, D. Lindel$\ddot{\rm{o}}$f$^1$, Lodi Rizzini$^{3,12}$,
\\ M. Macr\'\i$^2$, N. Madsen$^{6}$, G. Manuzio$^{2,7}$, M. Marchesotti$^5$,
\\ P. Montagna$^{3,4}$, H. Pruys$^1$, C. Regenfus$^1$, P. Riedler$^5$,
\\ A. Rotondi$^{3,4}$, G. Rouleau$^5$, G. Testera$^2$, A. Variola$^2$,
\\ L. Venturelli$^{12}$, D. P. van der Werf$^9$ and  N. Zurlo$^{12}$}


\address{$^1$Physik-Institut, Universit$\ddot{a}$t of Z$\ddot{u}$rich, CH-8057 Z$\ddot{u}$rich, Switzerland
\\ $^2$Istituto Nazionale di Fisica Nucleare, Sezione di Genova, 16146 Genova, Italy
\\ $^3$Istituto Nazionale di Fisica Nucleare, Sezione di Pavia, 27100 Pavia, Italy
\\ $^4$Dipartimento di Fisica Nucleare e Teorica, Universit\`{a} di Pavia, 27100 Pavia, Italy
\\ $^5$EP Division, CERN, CH-1211 Geneva 23, Switzerland
\\ $^6$Department of Physics and Astronomy, University of Aarhus, DK-8000 Aarhus C, Denmark
\\ $^7$Dipartimento di Fisica, Universit\`{a} di Genova, 16146 Genova, Italy
\\ $^8$Instituto de Fisica, Universidade Federal do Rio de Janeiro, Rio de Janeiro 21945-970, Brasil
\\ $^9$Department of Physics, University of Wales Swansea, Swansea SA2 8PP, UK
\\ $^{10}$Atomic Physics Laboratory, RIKEN, Saitama 351-0198, Japan
\\ $^{11}$Department of Physics, University of Tokyo, Tokyo 113-0033, Japan
\\ $^{12}$Dipartimento di Chimica e Fisica per l'Ingegneria e per i Materiali, Universit\`{a} di Brescia, 25123 Brescia, Italy}






\maketitle
\abstracts{In 2002, the ATHENA collaboration reported the creation and detection of cold ($\sim$15 K) antihydrogen atoms \cite{athenaNature}. The observation was based on the complete reconstruction of antihydrogen annihilations, simultaneous and spatially correlated annihilations of an antiproton and a positron. Annihilation byproducts are measured with a cylindrically symmetric detector system consisting of two layers of double sided Si-micro-strip modules that are surrounded by 16 rows of 12 pure CsI crystals \crystSize. This paper gives a brief overview of the experiment, the detector system, and event reconstruction.}

\section{Introduction}
Anithydrogen ({\hbar}) experiments are aimed at precisely testing CPT invariance and measuring the gravitational force between matter and antimatter with the simplest bound state of antiparticles. Many experimental advances towards accomplishing these goals have been made \cite{athenaNature,LEARanti,Fermianti,atrap}, in fact a little more than 1 year ago the ATHENA \cite{athenaNature} and ATRAP \cite{atrap} collaborations reported the production and detection of cold {\hbar}. Both of these collaborations employ similar techniques to create cold \hbar, though use different methods to detect the produced {\hbar}. This article begins with a short overview of the procedure for creating cold {\hbar}, though mainly focuses on the ATHENA detector system and the event reconstruction.

\section{Creating cold \hbar}
Cold {\hbar} has been produced by combining antiprotons ({\pbar}) with a cold ($\sim$ 15 K) positron (\pos) plasma ($\rho\sim$10$^{8}$ cm$^{-3}$). CERN's Antiproton decelerator  (AD) delivers {\pbar}s to the 4 AD experiments: ACE, ASACUSA, ATHENA and  ATRAP. ATHENA catches and cools a small fraction ($\sim$ 0.1\%) of the {\pbar} beam (n$_{\bar{p}}$ $\approx$ 10$^{7}$) from an AD spill ($\sim$ 100 s). Prior to capture, {\pbar}s travel through a thin degrader foil to further reduce their energy. From the energy distribution of {\pbar}s after the degrader foil, those in the tail on the low energy side are reflected by a $-$5 kV potential wall and then axially contained with another $-$5 kV wall that is upstream from the reflecting wall and raised after the {\pbar}s pass. Radial confinement near the axis is achieved with a 3 T magnetic field that is parallel to the electric field produced by the catching trap electrodes. In the catching trap {\pbar}s cool to the ambient temperature ($\sim$ 15 K) of the surrounding environment via Coulomb interactions with previously trapped electrons, which cool via synchrotron radiation. In parallel with the {\pbar} catching and cooling process, about 10$^8$ {\pos}s from a {\na} source are collected in a positron accumulator. The two antiparticle species are then transferred to and combined in a nested Penning trap \cite{Penning}, in which outer negative potential wells confine the {\pbar}s while an inner positive potential well confines the {\pos}s. In the nested Penning trap {\pbar}s loose energy as they traverse the inner {\pos} plasma. With some probability, {\pbar}s individually combine with {\pos}s to form {\hbar}, via three body recombination or radiative recombination. The neutral {\hbar} atoms travel freely though the charged particle confining electromagnetic fields and annihilate on the Au plated Al electrodes of the trap. This annihilation process may be viewed as two separate process, correlated both spatially and in time: an positron--electron annihilation usually \pos\ele$\rightarrow\gamma\gamma$; and  an antiproton--nucleon annihilation typically {\pbarn}$\rightarrow$5$\pi$, on average 3 charged and 2 neutral pions\cite{pbarClaude,NNbar_annihs}. Surrounding the {\hbar} formation region, the nested Penning trap, is a cylindrically symmetric detector system which has been designed to identify {\hbar}--annihilations. A more detailed description of the catching, cooling and combining processes may be found in \cite{AthenaNIM}.

\section{ATHENA's {\hbar} detector}
\begin{figure}[ht]
\centerline{\epsfxsize=3in\epsfbox{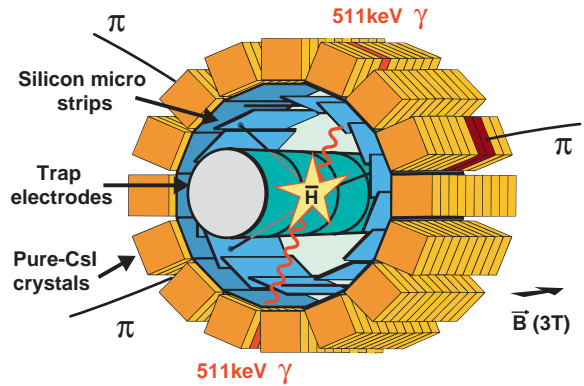}}   
\caption{Diagram of the ATHENA {\hbar} detector.\label{fig::detector}}
\vspace{-.2in}
\end{figure}

ATHENA's antihydrogen detector\cite{detectorPaper}, Fig. \ref{fig::detector}, is composed of two layers of silicon micro--strip/pad modules and a barrel of pure CsI crystals. The detector was designed to reconstruct charged particles from the {\pbarn}--annihilation with information from the two layers of silicon micro--strip/pad modules, and 511 keV photons from  {\pos\ele}--annihilations with the CsI crystals. Both the silicon layers and crystals are situated in a radially compact region, $\Delta$r $<$ 3 cm, spatially constrained by the inner trap electrodes and outer magnet. Environmental conditions is this region require the detector to operate in a 3 T magnetic field and at a temperature of 130 K.

\subsection{Silicon module layers and charged particle trajectory reconstruction}
\begin{figure}[ht]
\centerline{\epsfxsize=4in\epsfbox{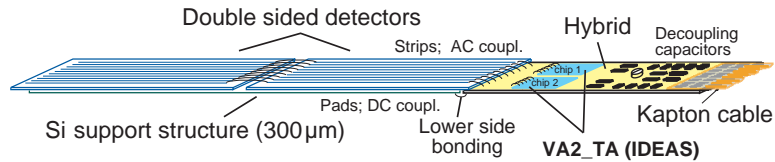}}   
\caption{Layout of a silicon strip/pad module.\label{fig::module}}
\end{figure}

Each of the two silicon layers consists of 16 identical modules oriented in cylindrical geometries (r$_{in} \approx$ 4 cm and r$_{out} \approx$ 5 cm). Two 8.16 $\times$ 1.9 cm$^2$ double side sensors, two 128 channel readout chips and a silicon support/bus compose each module, as shown in Fig. \ref{fig::module}. On the p-side of a sensor 128 strips are read out with a pitch of 139 $\mu$m. While the n--side is divided into 64 pads, 1.25 $\times$ 18 mm$^2$. In order to match the thermal expansion coefficients between the sensors and the support over a wide range of operating temperatures (80 -- 300 K), the sensors are glued to a silicon support structure/bus. The 128 line bus transfers signals from the pads to the front end electronics. The strips on the 2 sensors are wire bonded in series, as shown in Fig. \ref{fig::module}. A total of 128 strip channels and 128 pad channels are fed into the two readout chips on the module. Each input channel on the readout chip contains a pre--amp, a fast shaper ($\sim$100 ns), a discriminator and a slow ($\sim$2 $\mu$s) shaper. Signals from the pre--amps feed both the slow and the fast shapers. The pulse height discriminators compare the fast shaped signals to a common trigger--threshold. The outputs of 128 discriminators are combined with a logical ``or'' into one trigger output, so each module provides one strip and one pad side trigger signal. Signals from the slow shapers are sampled in the chip and multiplexed to a single analog output. These analog signals are transferred through external repeater cards to 10--bit VME ADC units (CAEN V550). See reference \cite{detectorPaper} for further details.

\begin{figure}[ht]
\centerline{\epsfxsize=4in\epsfbox{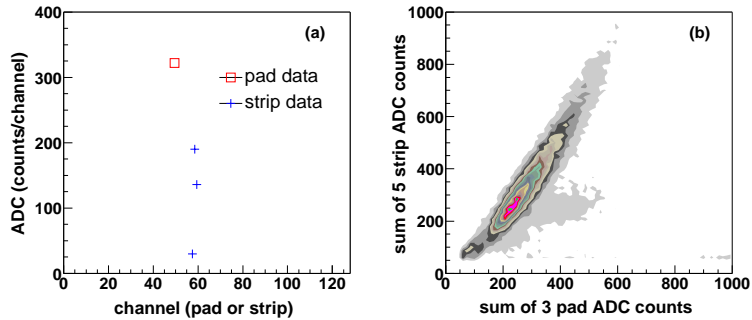}}   
\caption{Correlation between ADC counts on the pad and strip sides of the sensors. (a) Distribution of ADC counts across the pad and strip channels of a module for a minimum--ionizing particle. (b) Relation between the charge collected on the pad and strip sides of the sensors; distribution of mainly charged pions from \pbarn--annihilations in \pbar\pos mixing data.\label{fig::pad_strip}}
\end{figure}

Clusters are found from ADC information of the pads and strips. Typically 3 adjacent strips and 1 pad provide signals over a threshold level. Fig. \ref{fig::pad_strip}a illustrates the response on the pads and the strips from a minimum--ionizing charged particle. The physical position of a cluster is decoded from the module number, and strips and pads that fire. Multiple clusters on one module are resolved by requiring the sum of ADC counts on the strip and pad sides to be correlated\cite{AthenaNIM}. This correlation between the sum of the ADCs on the pad and strip sides is shown in Fig. \ref{fig::pad_strip}b. 


Trajectories of charged particles from {\pbarn}--annihilations are reconstructed from the cluster information of one inner and outer module. With the positions of only two clusters, the helical trajectories of independent charged particles are not uniquely defined. For this reason straight line fits are used to describe the trajectories. The position of the {\pbarn}--annihilation is found by minimizing the $\chi^2$ of functions that describe the crossing position of multiple trajectories in various planes\cite{AthenaNIM}. This method yields a vertex resolution of $\sigma\approx$ 4 mm. A more detailed description of cluster, track and vertex reconstruction may be found in reference \cite{AthenaNIM}.

\section{Pure CsI crystals and selecting 511 keV photon candidates}
Pure CsI crystals, read by avalanche photo diodes (APD), are used to identify 511 keV photon candidates, stemming from {\pos\ele}--annihilations.  Pure CsI crystals have the property that the light yield increases as the temperature decreases \cite{CsiTempDep,BornNIM}. This is an advantage in the ATHENA apparatus where T = 130 K.

\begin{figure}[ht]
\vspace{-.2in}
\centerline{\epsfxsize=3.4in\epsfbox{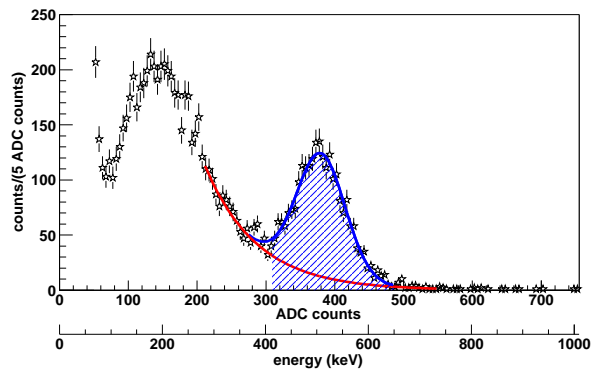}}   
\caption{Distribution of ADC counts of a crystal, for dedicated {\pos} runs. The 511 keV photon peak is fit with a Gaussian function plus an exponential background function. The shaded region indicates the acceptance window for 511 keV photons. \label{fig::crystAdcEn}}
\end{figure}

The barrel of crystals (r $\approx$ 5 cm, l = 16.2 cm) is arranged in 16 rows of 12 crystals \crystSize. The APDs of each row are readout with the a single chip (the same chip is used for the readout of the silicons sensors). Also as for the readout of the silicon modules, the slow shaped analog signals are transferred through external repeater cards to 10--bit VME ADC units (CAEN V550). These ADC values are linearly correlated to the energy deposited in the crystals (below $\approx$ 2 MeV). Crystal calibrations are performed on data from {\pos} only runs, in which the charged particle background is low and 511 keV photons are abundant. The 511 keV photon conversion peak in the ADC spectrum of each crystal is fit with a Gaussian function to determine the ADC value that corresponds to an energy of 511 keV, the resolution (FWHM $\approx$ 24\%) and the window for selecting 511 keV candidates ($-$2 to 3 $\sigma$), as shown in Fig. \ref{fig::crystAdcEn}.

\vspace{-.07in}
\section{Signatures of {\hbar} production}
\begin{figure}[ht]
\vspace{-.1in}
\centerline{\epsfxsize=2.35in\epsfbox{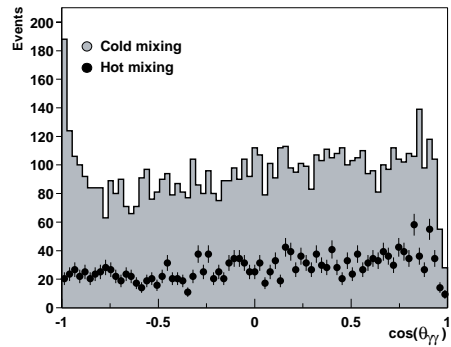}}
\caption{Distributions of the opening angle between two 511 keV photon candidates at the position of the reconstructed {\pbarn}--annihilation vertex. Cold mixing and hot mixing data sets refer to {\pos} plasma temperatures of approximately 15 K and 3500 K, respectively.\label{fig::cos}}
\end{figure}

ATHENA has shown the capability to fully reconstruct {\hbar}--annihilations, golden events, with information of the {\pbarn}--annihilation vertex from the silicon modules and 511 keV photon candidates from the crystals. A golden event is defined as an event with two 511 keV photon candidates emerging back--to--back from a reconstructed {\pbarn}--annihilation vertex. One additional constraint on the topology is that photon candidates are only considered in crystals which do not lie in the path of a charged particle or in the eight neighboring crystals. Golden events are identified with the distribution of the opening angle of 511 keV photons measured from the {\pbarn}--annihilation vertex. The enhancement of counts in the opening angle distribution near $\cos\left(\theta_{\gamma\gamma}\right)$ = $-$1 (back--to--back) statistically signifies the presence of golden events, as shown in Fig. \ref{fig::cos}. One consequence of this compelling signature of {\hbar} formation is its low reconstruction efficiency ($\sim$0.25\% for cos($\theta_{\gamma\gamma}$) $<$ $-$0.95) that is mainly due to the detection efficiency of a photon in the crystals ($\sim$20\%). However, it should be noted that it is not necessary to fully reconstruct events to study the properties of {\hbar} production. Other signatures which have a higher efficiencies, though a lower purity, are also utilized. In reference \cite{highRate} it has been shown that 65\% of all reconstructed {\pbarn}--annihilation vertices in cold mixing data are associated with {\hbar} atoms.

Current investigations which use a convolution of signatures to study the properties of {\hbar} production include the temperature dependence on the {\pos} plasma\cite{germanosTemp} and the n--states populated during {\hbar} formation. These and other ongoing measurements will improve our understanding of cold {\hbar} production, and ultimately aid future experiments that test CPT and measure the gravitational acceleration between matter and antimatter.

\end{document}